# A Trust-Based Detection Algorithm of Selfish Packet Dropping Nodes in a Peer-to-Peer Wireless Mesh Network

Jaydip Sen

Innovation Lab, Tata Consultancy Services Ltd,
Bengal Intelligent Pak, Salt Lake Electronics Complex, Kolkata - 700091, India
Jaydip.Sen@tcs.com

**Abstract.** Wireless mesh networks (WMNs) are evolving as a key technology for next-generation wireless networks showing raid progress and numerous applications. These networks have the potential to provide robust and high-throughput data delivery to wireless users. In a WMN, high speed routers equipped with advanced antennas, communicate with each other in a multi-hop fashion over wireless channels and form a broadband backhaul. However, the throughput of a WMN may be severely degraded due to presence of some selfish routers that avoid forwarding packets for other nodes even as they send their own traffic through the network. This paper presents an algorithm for detection of selfish nodes in a WMN that uses statistical theory of inference for reliable clustering of the nodes based on local observations. Simulation results show that the algorithm has a high detection rate and a low false positive rate.

## 1 Introduction

Wireless mesh networking has emerged as a promising concept to meet the challenges in next-generation networks such as providing flexible, adaptive, and reconfigurable architecture while offering cost-effective solutions to the service providers. Unlike traditional Wi-Fi networks, with each *access point* (AP) connected to the wired network, in WMNs only a subset of the APs are required to be connected to the wired network. The APs that are connected to the wired network are called the *Internet gateways* (IGWs), while the others are called the *mesh routers* (MRs). The MRs are connected to the IGWs using multi-hop communication. In a community-based WMN, a group of MRs managed by different operators form an access network to provide last-mile connectivity to the Internet. As with any end-user supported infrastructure, cooperative behavior in these networks cannot be assumed a priori. Preserving scarce access bandwidth and power, as well as security concerns may induce some selfish users to avoid forwarding data for other nodes. The selfish MRs degrade the routing performance in WMN by decreasing the network throughput [1].

To enforce cooperation among nodes and detect selfish nodes in ad hoc wireless networks, various collaboration schemes have been proposed in the literature [2]. Majority of these proposals are based on trust and reputation frameworks which attempts to identify misbehaving nodes by suitable decision making algorithms. To address the issue of selfish nodes in a WMN, this paper presents a scheme that uses local observations in the nodes for detecting node misbehavior. The scheme is applicable for on-demand routing protocols like AODV, and uses statistical theory of inference and clustering techniques to make a robust and reliable classification of the nodes based on their packet forwarding activities. It also introduces some additional fields in the packet header for AODV protocol so that detection accuracy is increased.

The rest of the paper is organized as follows. Section 2 presents some related work. Section 3 gives a brief background of the AODV protocol and a finite state machine model of the local observations of a node. The proposed scheme is described in Section 4. Section 5 presents simulations results, and Section 6 concludes the paper while identifying some potential future work.

## 2  Related Work

The concept of neighborhood monitoring to check the activities of other nodes has been proposed by researchers in the context of wireless ad hoc networks. The idea of watchdog mechanism to monitor neighbors was first proposed by Marti et al. [3]. Buchegger et al. have proposed the CONFIDANT protocol that assigns a rating for every node in an ad hoc network based on watchdog and second-hand rating information gathered from other nodes [4]. Mahajan et al. have proposed a mechanism named CATCH [5], which consists of two modules: (i) *anonymous challenge message* (ACM) and (ii) *anonymous neighbor verification* (ANV). First, an ACM message from an unknown sender is sent to all its neighbors. In the ANV phase, a tester node sends cryptographic hash of a random token for rebroadcast and also records other hashes sent by others. The tester node releases the secret token to another node which successfully authenticates itself. Vigna et al. have proposed an approach to detect intrusions in AODV that works on stateful signature-based analysis of the observed traffic [6]. Pirzada et al. have described a model of building trust relationship between nodes in an ad hoc network [7]. Conti et al. have proposed a scheme in which a node exploits its local knowledge to estimate the reliability of a path [8]. Unlike the conventional method of denying selfish users, it provides a degraded service to these nodes by selective slow packet forwarding. Santhanam et al. have presented a mechanism to judge the behavior of a node based on observed traffic reports submitted to local sink agents dispersed throughout the network [9]. The sink nodes apply a set of forwarding rules to isolate a selfish node based on the number of times it is caught in selfish acts. Tseng et al. have applied techniques based on finite state machines to detect misbehaving nodes in the AODV routing protocol [10]. Yang et al. have described the SCAN protocol that addresses two issues: (i) routing (control packets) misbehavior, and (ii) forwarding (data packets) misbehavior [11].

The proposed mechanism in this paper relies on local observation of each node in a WMN. Based on the local information in each node and using a finite state machine model of the AODV protocol, a robust statistical theory of estimation is applied to identify selfish nodes in the network. The scheme is a modification of the protocol proposed in [12]. The objective of the proposed mechanism is to achieve higher detection efficiency by exploiting the information in some additional fields in the packet header in AODV routing. The algorithm is discussed in Section 4.

## 3 AODV and Modeling of the State Machine

*Ad hoc on-demand distance vector* (AODV) routing protocol uses an on-demand approach for finding routes to a destination node. The source node floods the *route request* (RREQ) packet in the network when a route is not available for the desired destination. It may obtain multiple routes to different destinations from a single RREQ. The RREQ carries the source identifier (*src_id*), the destination identifier (*dest_id*), the source sequence number (*src_seq_num*), the destination sequence number (*dest_seq_num*), the broadcast identifier (*bcast_id*), and the *time to live* (TTL). When an intermediate node receives a RREQ, it either forwards the request further or prepares a *route reply* (RREP) if it has a valid route to the destination. Every intermediate node, while forwarding a RREQ, enters the previous node's address and its *bcast_id*. A timer is used to delete this entry in case a RREP is not received before the timer expires. When a node receives a RREP packet, information of the previous node from which the packet was received is also stored, so that data packets may be routed to that node as the next hop towards the destination.

It is clear that AODV depends heavily on cooperation among the nodes. A selfish node can easily manipulate it to minimize its chances of being included on routes for which it is not the source or the destination. The proposed mechanism detects selfish nodes in a WMN so that they may be isolated from the network. In the following subsection, the finite state machine model of the protocol is presented.

### 3.1 Finite State Machine Model

In the proposed mechanism, the messages corresponding to a RREQ flooding and the unicast RREP is referred to as a *message unit*. It is clear that no node in the network can observe all the transmission in a message unit. The subset of a message unit that a node can observe is referred to as the *local message unit* (LMU). The LMU for a particular node consists of the messages transmitted by the node and its neighbors, and the messages overheard by the node. The selfish node detection is done based on data collected by each node from its observed LMUs. For each message transmission in an LMU, a node maintains a record of its sender, and the receiver, and the neighbor nodes that receive the RREQ broadcast sent by the node itself.

The finite state machine shown in Fig. 1 depicts various states in which a node may exist for each LMU [12]. The states corresponding to the numbers mentioned in Fig.1 are listed in Table 1. The final states are *shaded*. Each message sent by a node causes a transition in each of its neighbor's finite state machine. The finite state machine in one neighbor gives only a local view of the activities of that node. It does not in any way, reflects the overall behavior of the node. The collaboration of each neighbor node makes it possible to get an accurate picture about the monitored node's behavior. In the rest of the paper, a node being monitored by its neighbors is referred to as a *monitored node*, and its neighbors are referred to as a *monitor node*. In the protocol, each node plays the dual role of a monitor node and a monitored node.

**Table 1.** The states of the finite state machine for a local message unit (LMU)

| State | Interpretation |
|---|---|
| 1: init | Initial phase; no RREQ is observed |
| 2: unexp RREP | Receipt of a RREP without RREQ observed |
| 3: rcvd RREQ | Receipt of a RREQ observed |
| 4: fwd RREQ | Broadcast of a RREQ observed |
| 5: timeout RREQ | Timeout after receipt of RREQ |
| 6: rcvd RREP | Receipt of a RREP observed |
| 7: LMU complete | Forwarding of a valid a RREP observed |
| 8: timeout RREP | Timeout after receipt of a RREP |

Each monitor node observes a series of interleaved LMUs for a routing session. Each LMU can be identified by the source-destination pair contained in a RREQ message. Let the $k^{th}$ LMU observed by a monitor node be denoted as $(s_k, d_k)$. The pair $(s_k, d_k)$ does not uniquely identify a LMU, because the source can issue multiple RREQs for the same destination. However, since the subsequent RREQs have some delays associated with them, it may be assumed that there is only one active LMU $(s_k, d_k)$ at any point of time. At the start of a routing session, a monitored node is at the state 1 in its finite state machine. As the monitor node(s) observes the behavior of the monitored node based on the LMUs, it records transitions form its initial state 1 to one of its possible final states -- 5, 7 and 8.

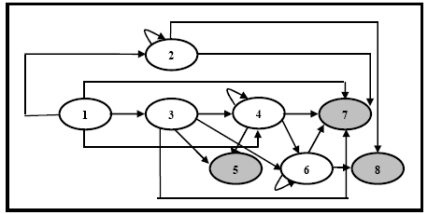

**Fig. 1.** Finite state machine of a monitored node

When a monitor node broadcasts a RREQ, it assumes that the monitored node has received it. The monitor node, therefore, records a state transition 1 → 3 for the monitored node's finite state machine. If a monitor node observes a monitored node to broadcast a RREQ, then a state transition of 3 → 4 is recorded if the RREQ message was previously sent by the monitor node to the monitored node; otherwise a transition of 1 → 4 is recorded since in this case, the RREQ was received by the monitored node from some other neighbor. The transition to a timeout state occurs when a monitor node finds no activity of the monitored node for the LMU before the expiry of a timer. When a monitor node observes a monitored node to forward a RREP, it records a transition to the final state – *LMU complete* (State No 7). At this state, the monitored node becomes a candidate for inclusion on a routing path. When the final state is reached, the state machine terminates and the state transitions are stored by each node for each neighbor. After sufficient number of events is collected, a statistical analysis is performed to detect the presence of any selfish nodes.

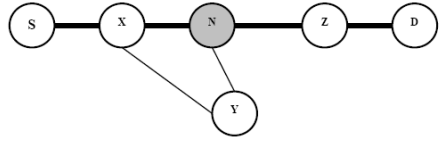

**Fig. 2.** An example local message unit (LMU) observed by node *N*

Fig. 2 depicts an example of LMU observed by the node *N* during the discovery of a route from the source node *S* to the destination node *D* indicated by bold lines. Table 2 shows the events observed by node *N* and the corresponding state transitions for each of its three neighbor nodes *X*, *Y* and *Z*.

**Table 2.** The state transitions of the neighbour nodes of node *N*

| Neighbor | Events | State changes |
|---|---|---|
| X | X broadcasts RREQ | 1 → 4 |
|   | N broadcasts RREQ | 4 → 4 |
|   | N sends RREP to X | 4 → 6 |
|   | X sends RREP to S (overheard) | 6 → 7 |
| Y | Y broadcasts RREQ | 1 → 4 |
|   | N broadcasts RREQ | 4 → 4 |
|   | Timeout | 4 → 5 |
| Z | N broadcasts RREQ | 1 → 3 |
|   | Z broadcasts RREQ | 3 → 4 |
|   | Z sends RREP to N | 4 → 7 |

## 4   The Proposed Algorithm

A monitoring node keeps track of state transitions in the finite state machine of a monitored node for each LMU. These sequences are represented as a transition matrix $T = [T_{ij}]$, where $T_{ij}$ is the number of times the transition $i \rightarrow j$ is found. The monitor node invokes a detection algorithm every *W* seconds using data from the most recent $D = d * W$ seconds of observations, where *d* is a small integer. The parameter *D*, the *detection window*, is such that it allows prompt punishment of the selfish nodes with a high level of accuracy. Section 4.1 discusses the features of the algorithm.

### 4.1   The Features of the Algorithm

While a transition matrix summarizes the local behavior of a monitored node, it is not possible to determine the selfish behavior of a node based only on its local transition probabilities. By comparing the transition matrices of a collection of nodes, one might be able to detect selfish nodes with higher confidence. For this reason, the proposed algorithm initially clusters the neighbors of a monitoring node and then classifies the clusters into selfish or cooperative nodes. The steps of the algorithm are: First, the clustering algorithm is made robust by the use of a statistical theory of inference-based approach that takes into account the pair-wise comparisons of the transition matrices of each pair of nodes. Second, for identification of the cluster that contains the selfish nodes, a measure, called *cooperation index*, for the nodes is computed. The cluster having its cooperation index less than a threshold value is assumed to contain the selfish nodes. Finally, a test is developed based on the *analysis of variance* (ANOVA) among the clusters to determine whether clustering is informative to the purpose of classification. In Section 4.2, the proposed algorithm is described.

### 4.2 The Detection Algorithm

In the proposed algorithm, a node is assumed to monitor the activities of its $R$ neighbors which are identified by their respective indices 1, 2,….R. Let $T^{(r)} = [f_{ij}^{(r)}]$ denote the observed transition matrix for the $r^{th}$ neighbor, where $[f_{ij}^{(r)}]$ is the number of transitions from state $i$ to state $j$ observed in the previous detection window. If $m$ is the number of states in the finite state machine in each node, the size of $T^{(r)}$ is $m \times m$. Let $T_i^{(r)} = [f_{i1}^{(r)},...f_{im}^{(r)}]$ denote the $i^{th}$ row of the transition matrix $T^{(r)}$, which shows the transitions out of state $i$ at the neighbor node $r$. If two neighbor nodes $r$ and $s$ have identical distributions corresponding to transitions from state $i$, then $T_i^{(r)} \equiv T_i^{(s)}$. To test the hypothesis $T_i^{(r)} \equiv T_i^{(s)}$ the Pearson's $\chi^2$ test is used as follows.

$$\chi^2(i) = \frac{\sum_{l\varepsilon(r,s)}\sum_{j=1}^{m}\left[f_{ij}^{(l)} - \bar{f}_{ij}^{(l)}\right]^2}{\bar{f}_{ij}^{(l)}} \quad (1)$$

$$\bar{f}_{ij}^{(l)} = F_{ij}^{(l)} \frac{f_{ij}^{(r)} + f_{ij}^{(s)}}{F_i^{(r)} + F_i^{(s)}}$$

$F_i^{(r)}$ and $F_i^{(s)}$ denote total number of transitions for state $i$ in $T^{(r)}$ and $T^{(s)}$ respectively. If the value of $\chi^2$ exceeds the value of $\chi^2_{m-1,\alpha}$, then the hypothesis $T_i^{(r)} \equiv T_i^{(s)}$ is rejected at confidence interval $\alpha$. If we write $K_i^{rs}$ for the event that $\chi^2_{(i)} > \chi^2_{m-1,\alpha}$, then the conditional probability $P(T_i^{(r)} \equiv T_i^{(s)} | B_i^{rs})$ can be taken as a reasonable estimator of the similarity between $r$ and $s$ with respect to the state $i$. In absence of any prior information, it is reasonable to assume that $r$ and $s$ have no similarity in state $i$ and the probability that the Pearson test rejects its hypothesis to be 0.5 [12]. In order to evaluate the similarity between $r$ and $s$ for all the $m$ states, (1) is applied to all rows of $T(r)$ and $T(s)$. This yields a vector $B^{(rs)} = [B_i^{(rs)}]$, $\{i = 1,\ldots m\}$. From the standard Markovian principle one can write:

$$L_{rs} = P(T^{(r)} \equiv T^{(s)} \mid B^{(rs)})$$

$$= \alpha^{S^{(rs)}} (1-\alpha)^{m-S^{(rs)}} \approx \alpha^{S^{(rs)}} \qquad (2)$$

$$\text{where} \quad S^{(rs)} = \sum_{i=1}^{m} B_i^{(rs)} \qquad (3)$$

The lower-order terms in the right hand side of (3) are ignored since α << 1. For small value of α, $L_{rs}$ monotonically decreases in $S^{(rs)}$, which, as evident from (3), represents the number of rejections of Pearson's hypothesis. Therefore, 1 - $L_{rs}$ may be taken as the measure of the dissimilarity between the neighbor nodes $r$ and $s$. In presence of noise, however, it is found that for two nodes $r$ and $s$ which have $L_{rs} \approx 1$, a third node $t$ may cause inconsistency such that $L_{rt} \neq L_{st}$. To avoid this, clusters are not formed on the basis of pair-wise dissimilarity. To compute dissimilarity between $r$ and $s$, the $L$ values for all neighbors are computed with respect to $r$ and $s$ separately, and (4) is applied:

$$d_{rs} = 1 - \frac{n_{rs}^2}{n_{r/s} * n_{s/r}} \qquad (4)$$

where, $n_{rs} = \sum_{t \neq r,s} \min(L_{rt}, L_{st})$, $n_{r/s} = \sum_{t \neq r,s}^{K} L_{rt}$, and $n_{s/r} = \sum_{t \neq r,s}^{K} L_{st}$.

The computation of $d_{rs}$ does not involve the pair-wise similarity index ($L_{rs}$) between nodes $r$ and $s$. It measures the degree of inconsistency in similarity between $r$ and $s$ with all their neighbors. Since in the computation of $d_{rs}$, contribution of each neighbor is considered, it is a robust indicator of dissimilarity between nodes [12]. For clustering, an *agglomerative hierarchical clustering* technique is used, in which each cluster is represented by all of the objects in the cluster, and the similarity between two clusters is measured by the similarity of the closest pair of data points belonging to different clusters. After the nodes are clustered into similar sets, they are further classified into three groups: (i) a set ($G$) of cooperative nodes, (ii) a set ($B$) of selfish nodes, and (iii) a set of nodes whose behavior could not be ascertained. The *cooperation score* ($C_r$) of a node is computed as [12]:

$$C_r = \frac{\sum_{i,j \in G}^{m} n_{ij}^{(r)}}{|G|} - \frac{\sum_{i,j \in B}^{m} n_{ij}^{(r)}}{|B|} \qquad (5)$$

To reduce the false positives (i.e. wrongly identification), an ANOVA test is applied that computes a probability $P_k$ of the random variation among the mean cooperation scores of $k$ clusters [12]. A small value of $P_k$ implies that the clusters actually represent differences in their behaviors. At each iteration, $k$ clusters are formed and $P_k$ is compared with a pre-defined level of significance β. If $P_k < β$, clusters reliably reflect the behavior of the nodes. The cluster with lowest mean cooperation score contains the selfish nodes. If $P_k > P_{k-1}$, the behavior of the nodes are not reflected in the clusters. In this case, all the nodes are classified as cooperative, and the next iteration of the algorithm is executed.

Even with the above statistical approach, there is still a possibility of misclassification. The probability of misclassification is further reduced by a new *cross-checking* mechanism that involves a minor modification in the AODV packet header. Two additional fields, *next_to_source* and *duplicate_flag* are inserted in a RREQ header to indicate respectively the next-hop address of the source, and whether the packet is a duplicate packet already broadcasted by some other nodes. In the RREP header, in addition to these two, another field called *next_to_destination* is added to indicate the node to which the packet is to be forwarded in the reverse path. With these additional fields, it is possible to detect every instance of selfish behavior in a wireless network, if the following conditions are satisfied: (i) no packet is lost due to interference, (ii) links are bi-directional, (iii) the nodes are stationary, and (iv) queuing delays are bounded [13]. The robust clustering and monitoring with additional fields substantially increase the detection as evident from the results.

## 5 Simulation Results

The protocol is evaluated with network simulator *ns-2* [14] in order to compare it with the algorithm in [12]. The simulation parameters are listed in Table 3.

**Table 3.** Simulation parameters

| Parameter | Value |
|---|---|
| Simulation area | 900 m * 900 m |
| Simulation duration | 1600 sec |
| No. of nodes in the network | 50 |
| MAC protocol | 802.11b |
| Routing protocol | AODV |
| Raw channel bandwidth | 11 Mbps |
| Traffic type | CBR UDP |
| Network traffic volume | 60 packets/sec |
| Packet size | 512 bytes |
| Time-out for RREQ broadcast | 0.5 sec |
| Time-out for receiving RREP | 3 sec |
| Pearson confidence (α) | 0.1 |
| Observation window (*W*) | 100 sec |
| Detection window (*D*) | 400 sec |
| Session arrival distribution | Poisson |
| Session duration distribution | Exponential |

At the start of the simulation, a fraction of nodes are chosen randomly as the selfish nodes. A selfish node adopts either of the two strategies: (i) dropping RREQs (DROP_REQ) and (ii) dropping RREPs (DROP_REP). In both cases, control packets are dropped with a constant probability. For DROP_REP, a selfish node always rebroadcasts RREQs even if it has a route in its cache. To evaluate the detection efficiency and speed, the packet dropping probability is varied from 1.0 to 0.1. The value of the parameter β is chosen as 0.4 to achieve the best tradeoff between detection rate and false positive rate.

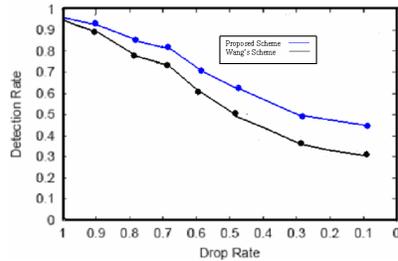 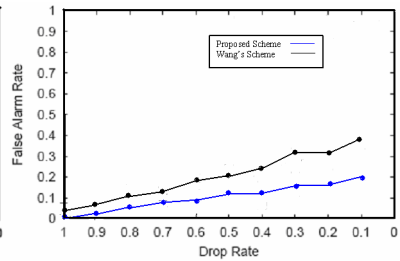

**Fig. 3.** The detection rate in DROP_REQ    **Fig. 4.** The false alarm rate in DROP_REQ

Fig. 3 and Fig. 4 represent respectively the detection rate and the false alarm rate when 50% nodes in the network are selfish and drop RREQs (i.e. DROP_REQ). The results are the average of 10 runs of the simulation. The algorithm performs better than Wang's algorithm since it doubly checks the detection results- from the clustering and from the routing header information to make more reliable detection.

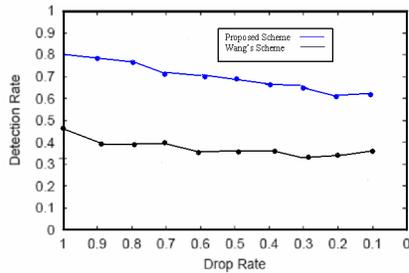 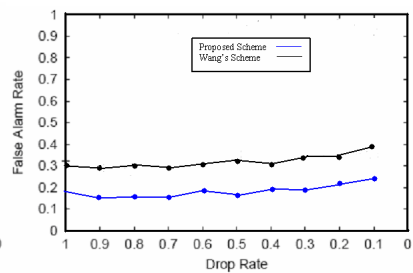

**Fig. 5.** The detection rate in DROP_REP    **Fig. 6.** The false alarm rate in DROP_REP

Fig. 5 and Fig. 6 show that the packet dropping (DROP_REP) has no impact on the detection rate and the false positive rate when 50% nodes in the network are selfish nodes. This difference in DROP_REQ and DROP_REP lies in the fact that while RREQ is a broadcast message sent by the source, the RREP is sent in a single path by the destination in a unicast manner. Since RREP involves only a few nodes, for majority of them the state machine will terminate in state 5, instead of states 7 and 8. It is evident that the proposed algorithm gives an average 80% increase in detection rate and 50 % reduction in false positives compared with the Wang's algorithm.

## 6   Conclusion and Future Work

Detection of selfish nodes is crucial in WMNs since these nodes don't forward packets for other nodes and degrade the performance of the networks. This paper has presented a statistical theory of inference-based clustering algorithm for detection of selfish nodes. Using the AODV protocol a finite state machine model is developed based on the local observations of each node. To increase the reliability of clustering, an ANOVA test and a new cross-checking mechanism are used. Simulation results show that the algorithm has high detection efficiency and reduced false alarm rates. Designing an efficient and secure routing algorithm that uses the output of the detection algorithm and avoids the selfish node constitutes a future plan of work.